\documentclass[12pt]{spieman}  % 12pt font required by SPIE;
\usepackage{amsmath,amsfonts,amssymb}
\usepackage{graphicx}
\usepackage{setspace}
\usepackage{tocloft}
\usepackage{siunitx}
\usepackage[english]{babel}
\usepackage{blindtext}
\usepackage{enumitem}
\usepackage{soul}
\usepackage{array}
\usepackage{multirow}
\usepackage{xfrac}
\usepackage{booktabs}
\usepackage{lineno}

\title{A Comprehensive Line-Spread Function Error Budget for the Off-Plane Grating Rocket Experiment}

\author[a,*]{Benjamin D. Donovan}
\author[a]{Randall L. McEntaffer}
\author[a]{James H. Tutt}
\author[a]{Bridget C. O'Meara}
\author[a]{Fabien Gris\'e}
\author[b]{William W. Zhang}
\author[b]{Michael P. Biskach}
\author[b]{Timo T. Saha}
\author[c]{Andrew D. Holland}
\author[c]{Daniel Evan}
\author[c]{Matthew R. Lewis}
\author[c]{Matthew R. Soman}
\author[d]{Karen Holland}
\author[d]{David Colebrook}
\author[d]{Fraser Cooper}
\author[d]{David Farn}

\affil[a]{The Pennsylvania State University, University Park, PA 16802}
\affil[b]{NASA Goddard Space Flight Center, Greenbelt, MD 20771}
\affil[c]{The Open University, Walton Hall, Milton Keynes, UK}
\affil[d]{XCAM Ltd., Northampton, UK}

\cftpagenumbersoff{figure}
\cftpagenumbersoff{table} 
\begin{document} 
\maketitle

% \linenumbers

\begin{abstract}
The Off-plane Grating Rocket Experiment (OGRE) is a soft X-ray grating spectrometer to be flown on a suborbital rocket. The payload is designed to obtain the highest-resolution soft X-ray spectrum of Capella to date with a resolution goal of R($\lambda/\Delta\lambda$) $>2000$ at select wavelengths in its 10--55 \si{\angstrom} bandpass of interest. The optical design of the spectrometer realizes a theoretical maximum resolution of $R\approx5000$, but this performance does not consider the finite performance of the individual spectrometer components, misalignments between components, and in-flight pointing errors. These errors all degrade the performance of the spectrometer from its theoretical maximum. A comprehensive line-spread function (LSF) error budget has been constructed for the OGRE spectrometer to identify contributions to the LSF, to determine how each of these affects the LSF, and to inform performance requirements and alignment tolerances for the spectrometer. In this document, the comprehensive LSF error budget for the OGRE spectrometer is presented, the resulting errors are validated via raytrace simulations, and the implications of these results are discussed.
\end{abstract}

% Include a list of up to six keywords after the abstract

% Include email contact information for corresponding author
{\noindent \footnotesize\textbf{*}Benjamin D. Donovan,  \linkable{bdonovan@psu.edu} }

% \begin{spacing}{2}   % use double spacing for rest of manuscript
\begin{spacing}{1}

\section{Introduction}
\label{sect:intro}  % \label{} allows reference to this section
The Off-plane Grating Rocket Experiment (OGRE) is a soft X-ray grating spectrometer that will be flown on a suborbital rocket. With a spectral resolution requirement of $R(\lambda/\Delta\lambda)>1500$ across its 10 -- 55 \si{\angstrom} bandpass of interest and a goal of $R>2000$ at select wavelengths in this same bandpass, OGRE will obtain the highest-resolution soft X-ray spectrum of Capella to date. This performance will enable OGRE to examine the spectrum of its target, Capella ($\alpha$ Auriga), in unprecedented detail. This detailed observation will permit existing line blends in the soft X-ray spectrum of Capella to be resolved, new and updated emission lines to be integrated into plasma spectral models, and more accurate plasma characteristics to be determined for this source \cite{Donovan:2019bb}. 

To achieve its performance goal of $R>2000$, OGRE will utilize three cutting-edge technologies: a mono-crystalline silicon X-ray optic assembly manufactured by NASA Goddard Space Flight Center (GSFC)\cite{zhang:2019bb}, six reflection grating modules developed by The Pennsylvania State University operated in the extreme off-plane mount \cite{McEntaffer:2013aa} and integrated into a grating assembly, and an array of four electron-multiplying CCDs (EM-CCDs) manufactured by e2v and integrated into a detector assembly by XCAM Ltd. and The Open University \cite{Lewis:2016aa}. 

\subsection{Optical Design of the OGRE Spectrometer} \label{sect:optical_design}

The OGRE spectrometer employs the traditional three-component X-ray grating spectrometer design consisting of an X-ray optic that focuses the incident light from Capella, a reflection grating array that diffracts this light into its component spectrum, and an array of detectors at the focal plane to sample the spectrum. 

The X-ray optic on board the OGRE spectrometer will be a mono-crystalline silicon X-ray optic assembly developed by NASA GSFC \cite{zhang:2019bb}. The optic assembly will consist of twelve Wolter I-type\cite{Wolter:1952aa} (paraboloid + hyperboloid) mirror shells with radii from $r_0=162.47-191.364$ mm and a common focal length of $Z_{0}=3500$ mm. Each of these shells will be constructed from individual silicon mirror segments that span 30\si{\degree} in azimuth. A more detailed description of this optic assembly can be found in Donovan et al. (2019) \cite{Donovan:2019bb}.

The OGRE spectrometer will utilize reflection gratings operated in the extreme off-plane mount to disperse the converging light from the OGRE optic assembly into its component spectrum. Gratings in the extreme off-plane mount are oriented quasi-parallel to the groove direction and at grazing incidence relative to the incident X-ray photons as depicted in Figure \ref{fig:opg_geometry}. Diffraction then follows the generalized grating equation:

\begin{figure}
\centering
\includegraphics[width=0.4\linewidth]{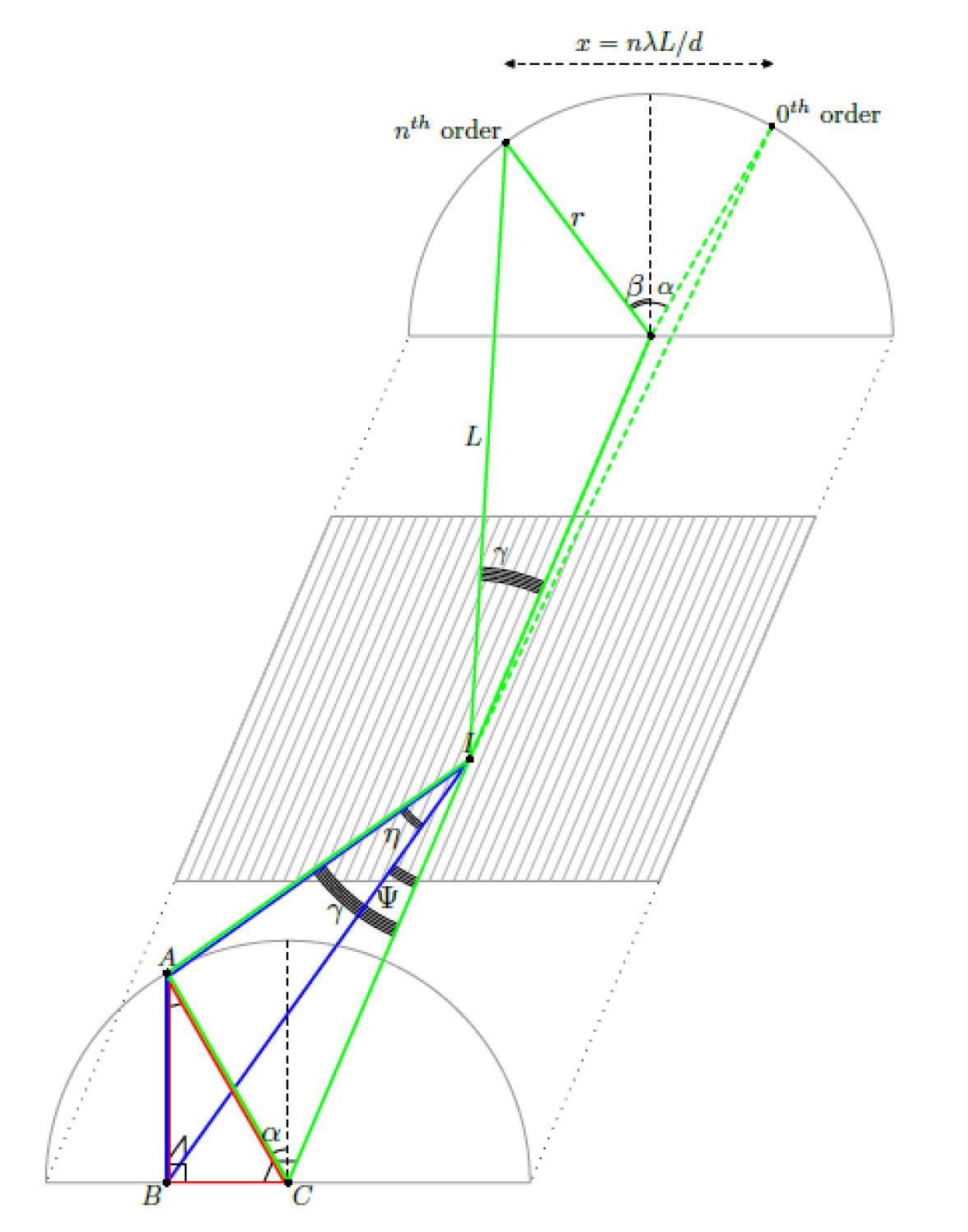}
\caption{The diffraction geometry of a reflection grating in the extreme off-plane mount \cite{Miles:2018aa, Cash:1991aa}. Light is incident from point A onto the grating surface at a graze angle $\eta$ and relative to the groove direction by an angle $\Psi$. Equivalently, this incidence geometry can be described in the spherical coordinate system by azimuth angle $\alpha$ and polar angle $\gamma$. Diffraction follows the generalized grating equation (Eq. 1) and light is diffracted a distance $L$ to azimuth angle $\beta$ on the focal plane. The dispersion distance between the $n=0$ reflection and $n$-th diffraction order on this focal plane is given by $x=n\lambda L/d$.}
\label{fig:opg_geometry}
\end{figure}

\begin{equation}\label{eq:1}
    \sin\alpha + \sin\beta = \frac{n\lambda}{d\sin\gamma},
\end{equation}

\noindent where $\alpha$ is the incident azimuthal angle, $\beta$ is the diffracted azimuthal angle, $\gamma$ is the polar angle between the incident light and the groove direction, $d$ is the groove period, $n$ is the diffraction order, and $\lambda$ is the wavelength of the light \cite{Cash:1983aa}. By differentiating Eq. \ref{eq:1} with respect to $\lambda$, it can be shown that:

\begin{equation}\label{eq:2}
    \frac{d\lambda}{dx} = \frac{10^7}{nLD} \frac{\si{\angstrom}}{\si{\milli\metre}},
\end{equation}

\noindent where $L$ is the distance a photon on the grating travels to the spectrometer focal plane, $D$ is the groove density ($\equiv 1/d$), and $x$ is the distance a photon is diffracted from the $n=0$ reflection ($=L\sin\gamma(\sin\beta+\sin\alpha)$). This equation shows that while the extreme off-plane mount diffracts its incident light conically, the spectral information is contained only in one dimension -- the dispersion direction ($x$ in Figure \ref{fig:opg_geometry}).

\begin{figure}
\centering
\includegraphics[width=0.8\linewidth]{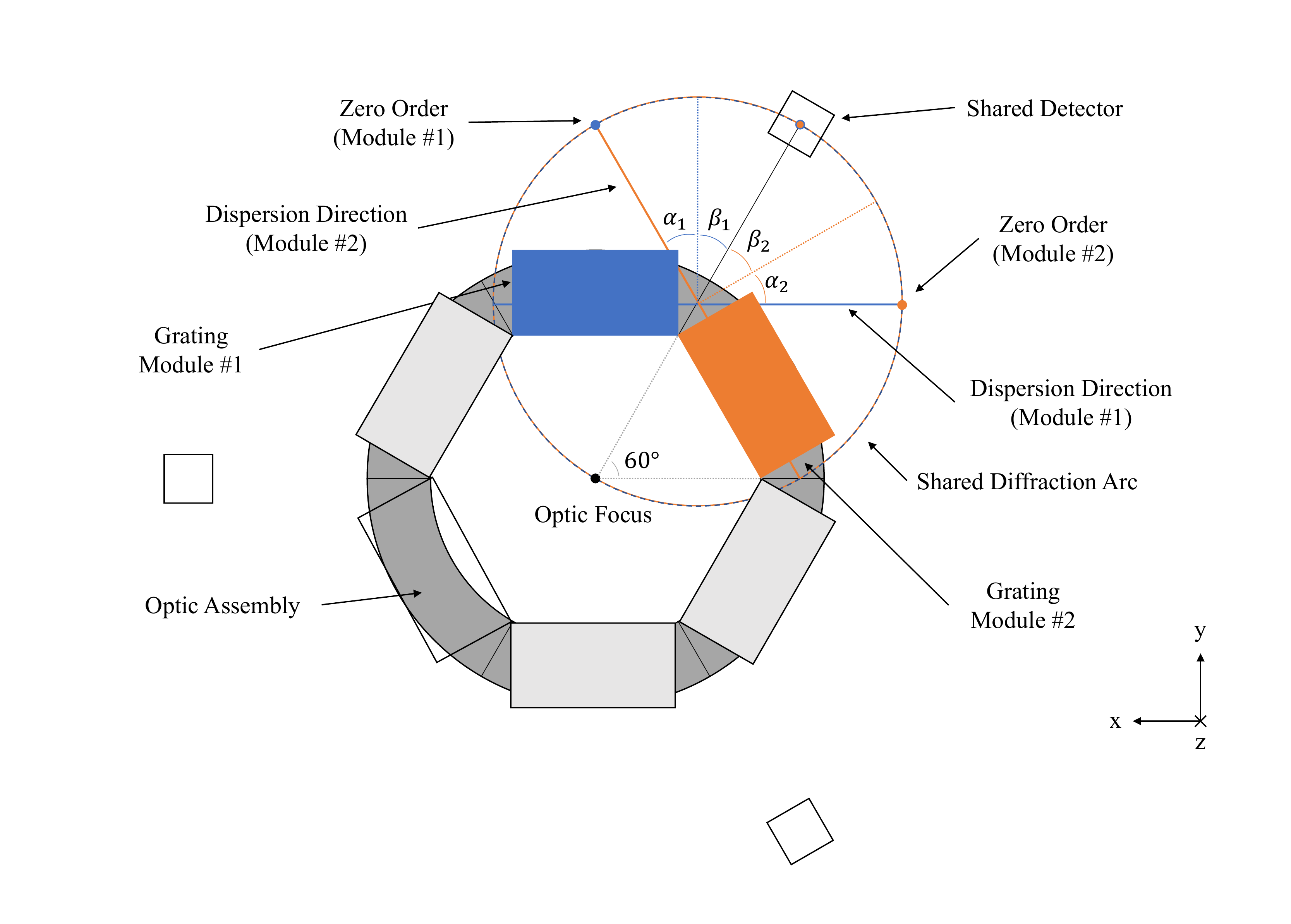}
\caption{Diffraction geometry for neighboring grating modules in the OGRE soft X-ray spectrometer \cite{Donovan:2019bb}. The optic assembly is divided into six 60\si{\degree} azimuthal sections. Behind each of these sections is a grating module containing 60 gratings. To maximize diffraction efficiency, each grating module operates in the Littrow mount which requires $\alpha=\beta=\delta$. For the OGRE spectrometer, $\alpha=\beta=\delta=30$\si{\degree} such that the neighboring grating modules diffract to the same location on the focal plane. This geometry is repeated two times to populate the entire 360\si{\degree} azimuthal span of optic with grating modules. Depicted in the bottom-right is the coordinate system for this schematic view with +\^{z} pointing into the page.}
\label{fig:ogre_grating_design}
\end{figure}

In the OGRE spectrometer, the full 360\si{\degree} azimuthal span of the optic assembly is divided into 60\si{\degree} azimuthal sections. Behind each of these azimuthal sections is a OGRE grating module containing 60 individual reflection gratings organized into two side-by-side grating stacks. The grating positions in each module are numerically optimized to realize maximum spectral resolution at the Fe XVII emission line ($\lambda=15.01$ \si{\angstrom}) -- the brightest line expected to be observed from Capella. Light from two neighboring grating modules is diffracted to the same location on the focal plane where the spectra are read out by a single detector. A schematic of this diffraction geometry can be seen in Figure \ref{fig:ogre_grating_design}. Spectral isolation on this detector will be attained through a combination of the detector's energy resolution and a slight offset ($\sim2-3$ mm) of the diffraction arcs relative to one another. This geometry is repeated two additional times around the optic assembly for a total of six grating modules diffracting to three separate spectral detectors (also shown in Figure \ref{fig:ogre_grating_design}). Further discussion of this geometry can be found in Donovan et al. (2019) \cite{Donovan:2019bb}.

The optical design of the OGRE spectrometer realizes a maximum spectral resolution of $R\approx5000$ at the Fe XVII emission line. This resolution is only attained though if all spectrometer components perform flawlessly, if these components are all perfectly aligned into the spectrometer, and if the payload remains oriented exactly towards its target. In reality, however, the measured performance of the gratings in the OGRE grating modules and the mirror segments in the OGRE optic assembly will deviate from their idealized performance. These performance errors will begin to degrade the diffracted line-spread function (LSF) of the spectrometer as observed on the focal plane. Furthermore, the spectrometer components will not be aligned perfectly to one other. These misalignments will further decrease the achievable performance of the spectrometer. Finally, the spectrometer will not remain perfectly pointed at Capella during its observation,  but will instead dither about its ideal pointing. This in-flight jitter will further degrade the observed LSF. These three sources of error will all conspire to cause the spectrometer to not achieve its theoretical spectral resolution of $R\approx5000$, but to achieve a performance below this level. 

% each component in the spectrometer does not perform at this level, they are not all aligned perfectly, and the payload does not remain oriented in this perfect manner. T

To understand how each misalignment and error contributes to the performance of the OGRE spectrometer, a comprehensive LSF error budget is required. This error budget identifies each contribution to the LSF, determines how each of these contributions affects the observed LSF, and then ultimately assigns requirements for each potential error or misalignment in the OGRE spectrometer so that the spectrometer can meet its overall performance requirement of $R>1500$. In this manuscript, the comprehensive LSF error budget for OGRE is described and implications resulting from this error budget are discussed.

\section{The Comprehensive LSF Error Budget}

The comprehensive LSF error budget considers potential misalignments and performance errors in the spectrometer. These misalignments and performance errors come from each of the three main components of the spectrometer: the OGRE optic assembly, the OGRE grating modules, and the OGRE detector assembly. An additional error arises from the in-flight pointing error of the payload. Each of these misalignments and performance errors modifies the observed LSF from the idealized LSF by increasing its extent in the dispersion and/or cross-dispersion direction. This behavior directly impacts the achievable spectral resolution and/or the effective area of the spectrometer. Additionally, misalignments can shift the centroid of the LSF on the detector. These shifts can move important spectral lines of Capella off of the detector, impacting the science return of the spectrometer. Thus, each potential performance error and misalignment must be analyzed to ensure the OGRE performance requirements are met.

\begin{figure}
\centering
\includegraphics[width=0.6\linewidth]{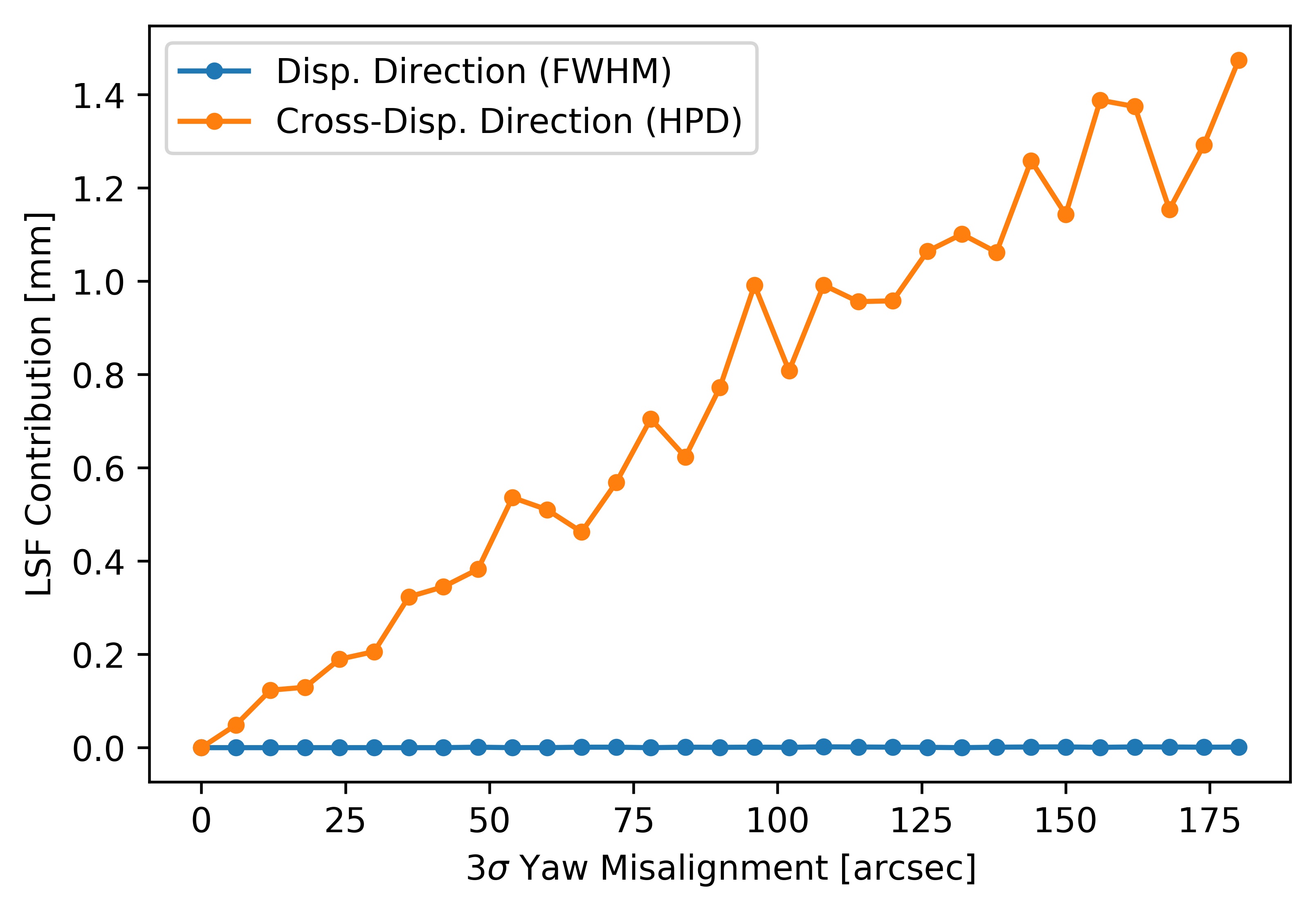}
\caption{Results from a raytrace simulation of the OGRE spectrometer showing the effect of a yaw misalignment of individual gratings in the grating module on the extent of the OGRE LSF in the dispersion (blue) and cross-dispersion (orange) directions. The contribution in the dispersion direction is measured as a full-width at half-maximum (FWHM), while the contribution in the cross-dispersion direction is measured as a half-power diameter (HPD). Here, it can been seen that a grating-level yaw misalignment at these simulated misalignment values does not contribute to the LSF extent in the dispersion direction, but it has a noticeable impact on the LSF extent in the cross-dispersion direction.}
\label{fig:misalignment-example}
\end{figure}

The construction of the LSF error budget begins with the consideration of a single contribution. For example, consider the yaw misalignment (rotation about \^{y}; degrees of freedom defined in Figure \ref{fig:ogre_grating_design}) of the 60 gratings within the two grating stacks that form Grating Module \#1 in Figure \ref{fig:ogre_grating_design}. Each of these gratings has a nominal yaw orientation within its stack, but each grating can be misaligned relative to this nominal orientation. To simulate the effect of this misalignment on the final LSF of the spectrometer, a custom raytrace model\footnote{Based upon PyXFocus: \linkable{https://github.com/rallured/PyXFocus}} of the OGRE spectrometer is utilized. A range of potential yaw misalignment values are simulated in this spectrometer model and the resulting LSF is measured for each. The results of these simulations are shown in Figure \ref{fig:misalignment-example}. In this figure, it can be seen that a yaw misalignment of the gratings within the two grating stacks affects the extent of the OGRE LSF in the cross-dispersion direction, but has no impact on the extent of the LSF in the dispersion direction. This agrees with what is expected from theory. Based on these results, a yaw alignment tolerance for the gratings within a stack would then be set (to be discussed in Section \ref{sect:grat_align}). Similar simulations would then be run for the remaining five degrees of freedom (\^{x}, \^{y}, \^{z}, pitch, and roll) of the gratings within their stacks, and requirements for these misalignments would be derived as well. Further simulations would then be performed for all remaining misalignments and performance errors within the spectrometer. 

Throughout the construction of the error budget, the performance requirements and goals are continually referenced to ensure that the spectrometer will meet these requirements and goals. For OGRE, the spectral resolution goal is $R>2000$ (resolution requirement: $R>1500$) which limits the extent of the LSF in the dispersion direction.  In the cross-dispersion direction, the entire LSF must remain within the planned window size of the spectral detector ($\sim6$ mm). While the LSF must remain within this window, movement of the LSF centroid in the cross-dispersion direction can be larger than this since the window can be moved on the detector. Total movement is limited to $\sim\pm12$ mm since the LSF must still remain on the detector ($\sim25$ mm x $25$ mm; e2v CCD207-40\cite{Lewis:2016aa}).

In the following subsections, individual contributions to the OGRE LSF error budget will be discussed. The discussion here is limited to a single 60\si{\degree} azimuthal section of the OGRE optic assembly, a single OGRE grating module placed behind this optic section, and a single spectral detector -- an OGRE spectrometer ``channel''. While this analysis is limited to a single channel, each channel is identical or mirrored with respect to the modeled channel, so the described analysis will apply to the remaining five spectrometer channels as well. Furthermore, this analysis is limited to the wavelength dispersing to the center of a spectral detector: $n\lambda=4.76$ \si{\nano\metre} ($x=98.2$ mm of dispersion).

\subsection{Optic Contributions}
 The development of the OGRE optic assembly is led by NASA GSFC \cite{zhang:2019bb}. Since this component is manufactured externally, a detailed error analysis of this component is beyond the scope of this error budget; however, similar error budgets for mono-crystalline optic assemblies have been developed for other X-ray missions such as \textit{Lynx} \cite{Zhang:2019aa}. The presented error budget only considers the final performance of a 60\si{\degree} azimuthal section of the OGRE optic assembly in the dispersion and cross-dispersion directions. 

\begin{table*}
\caption{Errors induced into the LSF of the OGRE spectrometer from a $60$\si{\degree} azimuthal section of the OGRE optic assembly. Since the optic will be treated as an assembled unit, the only errors to consider are its performance in the dispersion direction (measured as a full-width at half-maximum; FWHM) and the cross-dispersion direction (measured as a half-power diameter; HPD).} 
\label{tab:optic_contributions} 
\begin{center}   
\begin{tabular}{p{0.1\textwidth}p{0.17\textwidth}>{\centering}p{0.11\textwidth}>{\centering}p{0.11\textwidth}>{\centering}p{0.17\textwidth}>{\centering\arraybackslash}p{0.17\textwidth}}
\toprule
\multirow{2}{*}{Error}&\multirow{2}{*}{DoF}&\multicolumn{2}{c}{Requirement ($3\sigma$)}&\multicolumn{2}{c}{LSF Impact}\\\cline{3-6}
&&\si{\micro}m ($\pm$)&arcsec ($\pm$)&Disp. [\si{\micro}m]&X-Disp. [\si{\micro}m]\\
\hline
\multirow{2}{*}{PSF}&Disp. Dir.&--&1.5&25.4&--\\
&Cross-Disp. Dir.&--&5.0&--&84.8\\\hline
\multicolumn{4}{l}{\textbf{RSS Total}}&25.4&84.8\\
\bottomrule
\end{tabular}
\end{center}
\end{table*} 
 
% \begin{table}[ht]
% \caption{Errors induced to the LSF of the OGRE spectrometer from a $60$\si{\degree} azimuthal section of the OGRE optic assembly. Since the optic will be treated as an assembled unit, the only errors to consider are its performance in both the dispersion direction (measured as a full-width at half-maximum; FWHM) and the cross-dispersion direction (measured as a half-power diameter; HPD).} 
% \label{tab:optic_contributions}
% \begin{center}       
% \begin{tabular}{|l|l|} %% this creates two columns
% %% |l|l| to left justify each column entry
% %% |c|c| to center each column entry
% %% use of \rule[]{}{} below opens up each row
% \hline
% \rule[-1ex]{0pt}{3.5ex}  Document entity & Brief description  \\
% \hline\hline
% \rule[-1ex]{0pt}{3.5ex}  Article title & 16 pt., bold, left justified  \\
% \hline
% \rule[-1ex]{0pt}{3.5ex}  Author names & 12 pt., bold, left justified   \\
% \hline
% \rule[-1ex]{0pt}{3.5ex}  Author affiliations & 10 pt., left justified   \\
% \hline
% \rule[-1ex]{0pt}{3.5ex}  Abstract & 10 pt.  \\
% \hline
% \rule[-1ex]{0pt}{3.5ex}  Keywords & 10 pt.  \\
% \hline
% \rule[-1ex]{0pt}{3.5ex}  Section heading & 12 pt., bold, left justified  \\
% \hline
% \rule[-1ex]{0pt}{3.5ex}  Subsection heading & 12 pt., italic, left justified  \\
% \hline
% \rule[-1ex]{0pt}{3.5ex}  Sub-subsection heading & 11 pt., italic, left justified  \\
% \hline
% \rule[-1ex]{0pt}{3.5ex}  Normal text & 12 pt. \\
% \hline
% \rule[-1ex]{0pt}{3.5ex}  Figure and table captions &  10 pt. \\
% \hline 
% \end{tabular}
% \end{center}
% \end{table} 

The performance requirements of the OGRE optic assembly are derived from the current performance of mono-crystalline optic segments. Single paraboloid-hyperboloid mirror pairs routinely achieve a point spread function (PSF) with half-power diameter (HPD) of $<2$ arcsec \cite{zhang:2019bb}. Conservative estimates suggest that this performance will degrade slightly to $\sim3-5$ arcsec HPD when all 288 individual segments are aligned together. In the dispersion direction, a $60$\si{\degree} azimuthal section of the optic assembly with this performance is expected to perform at $<1.5$ arcsec full-width at half-maximum (FWHM). Thus, the OGRE optic performance requirements for a 60\si{\degree} azimuthal section of the optic assembly that feeds a single grating module have been set at $<1.5$ arcsec FWHM in the dispersion direction and $<5$ arcsec HPD in the cross-dispersion direction. At a focal length of $Z_{0}=3500$\,mm -- the focal length of the OGRE optic assembly -- these dispersion and cross-dispersion requirements correspond to $<25.4$ \si{\micro}m FWHM and $<84.8$ \si{\micro}m HPD, respectively. These errors are summarized in Table \ref{tab:optic_contributions}. 

The finite optic performance is thus the first contribution impacting the LSF of the OGRE spectrometer. If the only contribution to the LSF is the PSF from this optic, the spectrometer will exactly reproduce the PSF at dispersion. With the dispersion limited to $x\sim98.2$ mm due to the size of the payload, this equates to a maximum achievable spectral resolution of $R(x/\Delta x)\approx3860$ for the system. While this performance is well beyond the OGRE spectral resolution goal of $R>2000$, there are many additional contributions in the system that also impact to the LSF of the OGRE spectrometer.

% The performance of the OGRE spectrometer has thus already degraded slightly due to the finite optic performance. However, there are many additional contributions to the LSF that must be considered which will degrade the performance of the spectrometer further.

\subsection{Grating Contributions}
Beyond the finite performance of the optic, additional grating-related errors are introduced into the observed LSF of grating spectrometers. In OGRE, these grating-related errors arise from five main sources: aberrations induced by the diffraction geometry, the finite performance of the individual gratings within the grating module, and three misalignments during the construction of the grating module and the alignment of this grating module to a 60\si{\degree} azimuthal section of the OGRE optic assembly. Each of these five error sources will be discussed in the following subsections. 

% A grating module is composed of two grating stacks which each contain 30 gratings. The top-level process of grating alignments starts with the alignment of the individual grating substrates themselves (grating-to-grating alignment).  The grating substrates will be stacked on top of one another to form a grating stack. Then, two of these grating stacks will be aligned together to form a grating module (stack-to-stack alignment). This grating module is then aligned to a 60\si{\degree} azimuthal section of the OGRE optic assembly (module-to-optic alignment) to form the forward assembly. In each of these alignment steps, misalignments will occur.

\subsubsection{Diffraction Geometry Aberration}

The first contribution to the OGRE LSF from the gratings in the spectrometer comes from the diffraction geometry itself. While the goal of a reflection grating spectrometer such as OGRE is to exactly reproduce the optic PSF at dispersion, this rarely happens in practice. Aberrations are induced into the system which blur the diffracted-order LSF and cause it to diverge from the optic PSF. These aberrations can arise from several sources, including diffraction-induced astigmatism and the sampling of a curved focal plane with a flat detector. For OGRE, the grating positions were numerically optimized such that these aberrations were eliminated in the dispersion direction at Fe XVII ($\lambda=15.01$ \si{\angstrom}). However, while these aberrations were eliminated in the dispersion direction, the extent of the LSF in the cross-dispersion direction grew slightly durin this numerical optimization exercise; the LSF error in the cross-dispersion direction was 60.5 \si{\micro}m HPD. A summary of this induced LSF error is listed in Table \ref{tab:grating_contributions} as ``Diff. Aberration''. 

% Since no additional errors were induced in the dispersion direction, the spectral resolution remains at $R\approx3860$.

\begin{table*}
\caption{Grating-induced errors to the observed LSF of the OGRE spectrometer, including aberrations induced by the diffraction geometry, the finite resolution limit of the individual gratings within a module, grating-level alignment, stack-level alignment, and module-to-optic alignment. Shown are the $3\sigma$ level (99.7\%) requirements for each error in all six degrees of freedom (DoF; if applicable) and the impact of the error in both the dispersion direction (measured as a full-width at half maximum [FWHM]) and the cross-dispersion direction (measured as a half-power diameter [HPD]). Only the maximum LSF impact values are reported for each error in this table.}
\label{tab:grating_contributions}
\begin{center}   
\begin{tabular}{p{0.17\textwidth}>{\centering}p{0.1\textwidth}>{\centering}p{0.13\textwidth}>{\centering}p{0.13\textwidth}>{\centering}p{0.15\textwidth}>{\centering\arraybackslash}p{0.15\textwidth}}
\toprule
\multirow{2}{*}{Error}&\multirow{2}{*}{DoF}&\multicolumn{2}{c}{Requirement ($3\sigma$)}&\multicolumn{2}{c}{LSF Impact}\\\cline{3-6}
&&\si{\micro}m ($\pm$)&arcsec ($\pm$)&Disp. [\si{\micro}m]&X-Disp. [\si{\micro}m]\\
\toprule
Diff. Aberration&--&--&--&--&60.5\\\hline
Grat. Res. Limit&--&\multicolumn{2}{c}{$\text{R}=\text{4500}$}&21.8&--\\\hline
\multirow{6}{2.5cm}{Grating Alignment Within Stack}&X&127&--&0.4&102.6\\
&Y&127&--&0.8&--\\
&Z&127&--&8.5&--\\
&Pitch (X)&--&30&--&489.9\\
&Yaw (Y)&--&30&0.6&282.2\\
&Roll (Z)&--&15&10.8&5.7\\\hline
\multirow{6}{2.5cm}{Stack Alignment Within Module}&X&127&--&1.0&179.4\\
&Y&254&--&--&551.9\\
&Z&127&--&6.7&21.4\\
&Pitch (X)&--&60&0.3&1910.8\\
&Yaw (Y)&--&60&3.6&1130.7\\
&Roll (Z)&--&30&31.0&62.6\\\hline
\multirow{6}{2.5cm}{Module Alignment To Optic Section}&X&500&--&1.0&10.4\\
&Y&500&--&2.0&10.9\\
&Z&1000&--&1.7&39.9\\
&Pitch (X)&--&60&1.0&9.2\\
&Yaw (Y)&--&60&2.1&13.8\\
&Roll (Z)&--&120&1.6&4.7\\\hline
\multicolumn{4}{l}{\textbf{RSS Total}}&41.2&2367.9\\
\bottomrule
\end{tabular}
\end{center}
\end{table*}

% \begin{table}
% \centering
% \caption{Grating-induced errors to the observed LSF of the OGRE spectrometer, including aberrations induced by the diffraction geometry itself, the finite resolution limit of the individual gratings, grating-level alignment, stack-level alignment, and module-to-optic alignment. Shown are the $3\sigma$ level (99.7\%) requirements for each error in all six degrees of freedom (DoF; if applicable) and the impact of the error in both the dispersion direction (measured as a full-width at half maximum [FWHM]) and the cross-dispersion direction (measured as a half-power diameter [HPD]). }
% \includegraphics[width=\linewidth]{grating-errors-rev10.pdf}
% \label{tab:grating_contributions}
% \end{table}

\subsubsection{Grating Performance}
The second contribution to the LSF from the gratings is the aberration due to the finite resolution of the gratings that form an OGRE grating module. The OGRE spectrometer is designed to utilize gratings with a radial groove profile and a groove period of 160 nm at a distance of 3300 mm from the hub of the converging grooves \cite{Donovan:2020aa}. However, the processes used to manufacture these gratings will introduce errors that cause the manufactured groove pattern to deviate slightly from the idealized groove pattern. This deviation is expected to be a random process; therefore, the groove period at any given location on a grating will have a Gaussian distribution about its nominal value. From Eq. \ref{eq:2}, this Gaussian groove period error will manifest on the focal plane as a Gaussian error in the dispersion direction. These Gaussian period errors do not impact the cross-dispersion extent. 

An OGRE grating prototype was tested for spectral resolution at the Max Planck Institute for Extraterrestrial Physics' PANTER X-ray Test Facility in an attempt to measure groove-induced aberrations. Results from this testing indicate that the OGRE grating prototype achieved a groove-induced spectral resolution of $R_g>4500$ at the $\sim94$\% confidence level \cite{Donovan:2020aa}. However, this achieved spectral resolution was found to be limited by the measurement limit of the assembled spectrometer. Thus, the true grating-induced aberrations of the OGRE grating prototype could not be measured in this test and could have been higher than this limit. Similar X-ray gratings produced via electron-beam lithography for synchrotron applications ($d=500$ nm) were measured to have groove period errors of $<0.01$ nm \cite{Voronov:2017, Deroo:2020aa}. If similar groove period errors can be achieved for OGRE gratings with a period of $d=160$ nm, these gratings could achieve $R(d/\Delta d)\gtrsim16,000$. However, since this limit has not been measured explicitly for OGRE-like periods, a conservative requirement of $R>4500$ ($3\sigma$) has been adopted for the grating-induced resolution in the LSF error budget. If each grating performs at this level, the impact on the observed LSF in the dispersion direction is 21.8 \si{\micro}m FWHM. This error is listed in Table \ref{tab:grating_contributions} as the ``Grat. Res. Limit''.

\begin{figure}
\centering
\includegraphics[width=0.98\linewidth]{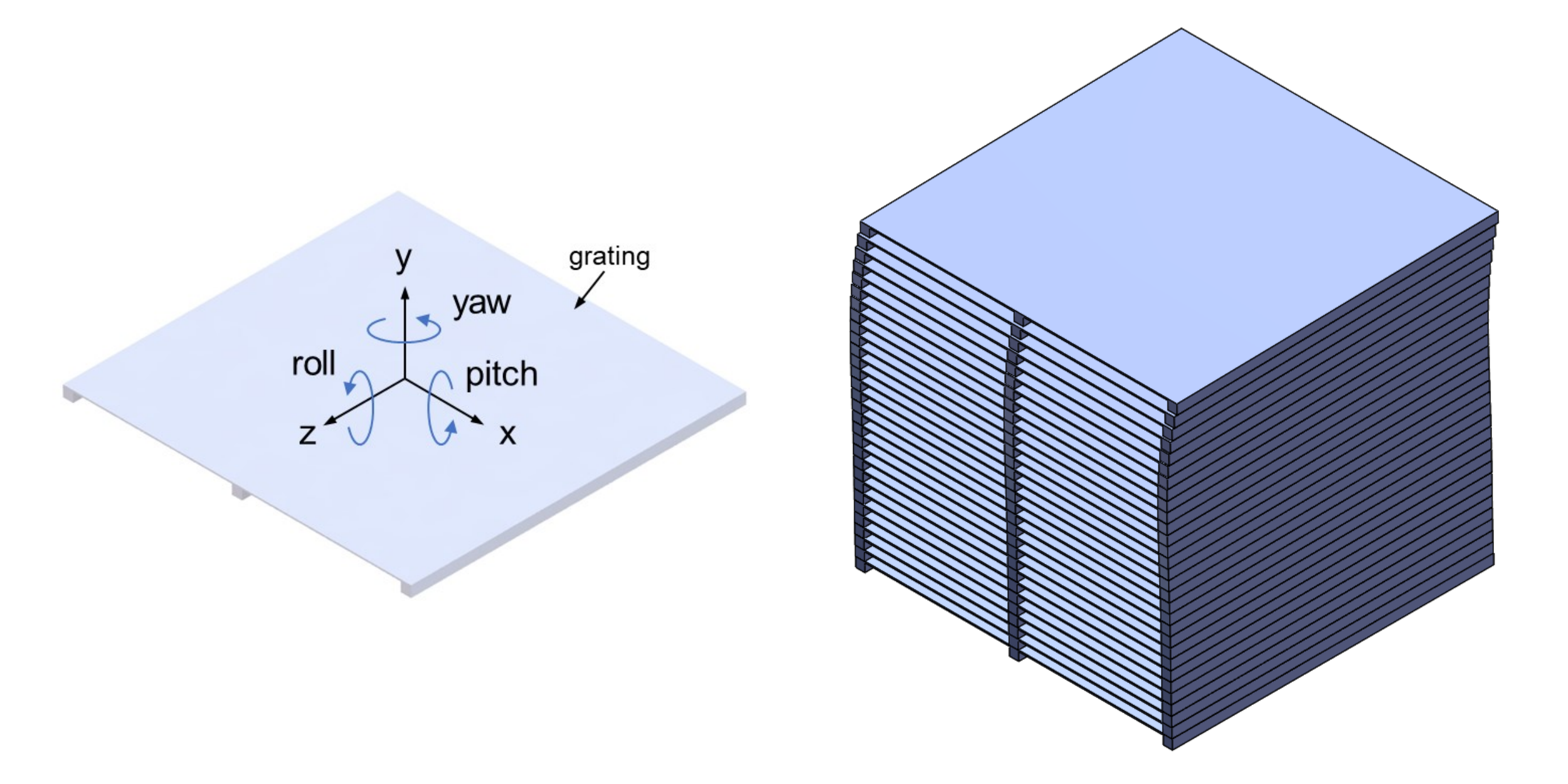}
\caption{\textit{Left} - A single reflection grating in the OGRE spectrometer \cite{OMeara:2019aa}. Each grating has three 2 mm wide ribs manufactured into the back of the substrate which allow light to pass through the grating stack while still maintaining the precisely manufactured wedge profile. Overlayed on this figure are the degrees of freedom (DoF) for each reflection grating. The grooves for each grating converge in the $-$\^{z} direction to the spectrometer focal plane. \textit{Right} - An OGRE grating stack containing 30 individual reflection gratings stacked on top of one another.}
\label{fig:grating-dof}
\end{figure}

\subsubsection{Grating Alignment} \label{sect:grat_align}
With the grating pattern manufactured, the construction of an OGRE grating module can begin. Individual OGRE gratings start as wedged silicon substrates. These substrates are manufactured such that their wedge angle replicates the fan angle between neighboring gratings in each stack as required by the OGRE optical design \cite{Donovan:2019bb}. Each substrate will then be imprinted with the OGRE grating pattern ($70$ mm x $70$ mm)  via substrate conformal imprint lithography \cite{Verschuuren:2017aa}. The grating pattern will then be precisely diced from the wedged substrate such that the sides of each grating are aligned relative to the grating pattern itself. In addition to the dicing process, the majority of the backside of each grating substrate will be removed leaving a face sheet with a thickness of $\sim0.3-0.5$ mm and three 2 mm wide x 70 mm long ``ribs'' (depicted in Figure \ref{fig:grating-dof}).  These ribs allow light to pass through each OGRE grating stack, but also maintain the precise wedge profile of each grating within the stack. The backside of each grating substrate is then etched to remove the stress introduced during this ``ribbing'' process. This completes the manufacture of a grating substrate \cite{Donovan:2019bb, OMeara:2019aa}. 

Once individual grating substrates have been manufactured, they are then stacked on top of each other to realize aligned grating stacks as depicted in Figure \ref{fig:grating-dof}. The wedged grating substrates themselves largely constrain the grating-level pitch, roll, and \^{y} alignment during this stacking process (degrees of freedom shown in Figure \ref{fig:grating-dof}). A precision robot will be used to achieve alignment in the remaining three degrees of freedom (\^{x}, \^{z}, and yaw) by referencing the sides of each grating substrate and by maintaining a precise global coordinate system during the stacking process. This alignment methodology is similar to that utilized for the silicon pore optic (SPO) technology developed by cosine Research B.V. \cite{Collon:2018aa} -- a collaborator on these alignment efforts. 

The optical design of the spectrometer gives the desired placement of these gratings within the two stacks. However, the manufacture of each grating and the assembly of the 60 gratings into the two grating stacks will result in placement errors of each grating relative to their designed placement. These misalignments introduce aberrations into the observed LSF which will affect the achievable performance of the OGRE spectrometer. The impact of a grating-level misalignment in each degree of freedom is discussed below.

\begin{itemize}[label={}]
    \item \textbf{\^{x}:} The grating pattern will be diced from the grating substrate to an accuracy of $<2$ \si{\micro}m over the 70\,mm length of the grating pattern. The stacking robot will then place this grating substrate into the stack. The precision of the stacking robot is $\lesssim30$ \si{\micro}m, but the interface between the robot and the grating substrate is currently unknown. In addition to the stacking robot itself, this interface is another source of error in the stacking process. Since this interface is currently unknown, the alignment tolerance in this degree of freedom has been set to $\pm127$ \si{\micro}m ($3\sigma$; standard machine tolerance). A misalignment at this level mainly impacts the cross-dispersion extent of the observed LSF with an contribution of $102.6$ \si{\micro}m HPD, but has a small effect on the dispersion direction as well ($0.4$ \si{\micro}m FWHM).
    \item \textbf{\^{y}:} Alignment in this degree of freedom is largely constrained by wedge manufacture. The manufacturer can easily meet standard machine tolerances in this degree of freedom, so this tolerance level has been adopted for this degree of freedom: $\pm127$ \si{\micro}m ($3\sigma$). A \^{y} misalignment at this level only has a slight contribution to the LSF extent in the dispersion direction: $0.8$ \si{\micro}m FWHM, but does not impact the LSF in the cross-dispersion direction.
    \item \textbf{\^{z}:} Just as with \^{x} alignment, alignment in this degree of freedom is achieved largely by the stacking robot. Since the exact interface between the robot and the grating substrate is unknown, this tolerance has been set to $\pm127$ \si{\micro}m ($3\sigma$; standard machine tolerance). A \^{z} misalignment at this level does not affect the observed LSF in the cross-dispersion direction, but has a significant impact on the dispersion direction: $8.5$ \si{\micro}m FWHM. 
    \item \textbf{Pitch (rotation about \^{x}):} A pitch misalignment of an individual grating relative to its nominal orientation acts to move the diffraction arc in the cross-dispersion direction. Misalignments of all 60 gratings in pitch will then increase the extent of the combined LSF in the cross-dispersion dimension. The wedge manufacturer can achieve a tolerance on the wedge angle of $\pm30$ arcsec ($3\sigma$), so this tolerance has been adopted as the grating-level pitch alignment tolerance. A pitch misalignment of each grating at this level increases the extent of the LSF in the cross-dispersion direction by $489.9$ \si{\micro}m HPD. 
    \item \textbf{Yaw (rotation about \^{y}):} A yaw misalignment of the gratings within the grating stacks will increase the extent of the LSF in the cross-dispersion direction. The dicing process has an accuracy of $<2$ \si{\micro}m over the 70 mm length of the grating such that the grating edges will be aligned to $<6$ arcsec relative to the grating pattern itself. To limit the cross-dispersion impact on the LSF, the yaw alignment tolerance is $\pm30$ arcsec ($3\sigma$). A yaw misalignment at this level will increase the extent of the LSF in the cross-dispersion direction by $282.2$ \si{\micro}m HPD and will slightly increase the extent in the dispersion direction by 0.6 \si{\micro}m FWHM.
    \item \textbf{Roll (rotation about \^{z}):} This degree of freedom is constrained both by wedge manufacture and the dicing process. The wedges can be manufactured such that the top and bottom surfaces are misaligned in roll by no more that $\pm15$ arcsec. The grating pattern will then be aligned to the wedge direction to within $\pm0.5$\si{\degree}. Once aligned, the grating pattern is then diced to within $<2$ \si{\micro}m over the 70 mm length of the grating. With the grating pattern aligned to within $\sim0.5$\si{\degree} of the wedge direction, there is no additional roll induced by a misalignment of the wedged substrate and the grating pattern. Therefore, the roll alignment tolerance is set to that which is achievable during wedge manufacture: $\pm15$ arcsec ($3\sigma$). A misalignment at this level increases the extent of the LSF in the dispersion direction by $10.8$ \si{\micro}m FWHM and slightly increases the extent in the cross-dispersion direction by $5.7$ \si{\micro}m HPD.
\end{itemize}

\noindent As mentioned previously, the interface between the stacking robot and the grating substrates is currently unknown. Therefore, there are some uncertainties in the achievable alignment of the gratings within stacks. While \^{x} and \^{z} alignment tolerances were purposefully set to standard machine tolerances to account for this unknown interface, the yaw tolerance is much tighter than can be achieved by standard machine tolerances. Further investigation will be needed to determine the interface between the grating substrates and the stacking robot, and if an additional constraint mechanism is needed. Precision pins are currently being investigated to serve as this additional constraint. Since the edges of the gratings will be aligned to $<6$ arcsec relative to the grating pattern itself from the dicing process, each grating side can reference two precision pins to constrain grating yaw within a stack. Additionally, while the error budget baselines the wedged grating alignment methodology presented here and in Donovan et al. (2019) \cite{Donovan:2019cc}, this error budget could easily be adapted for other grating alignment methods if needed. 

\subsubsection{Stack Alignment}
Once the grating stacks have been constructed, they must be aligned into the OGRE spectrometer. Rather than directly aligning the two stacks relative to the optic, the stacks will first be integrated into a grating module. This greatly eases spectrometer integration, but allows for additional misalignments to be introduced into the spectrometer and therefore additional aberrations to be introduced into the observed LSF. 

\begin{figure}
\centering
\includegraphics[width=0.6\linewidth]{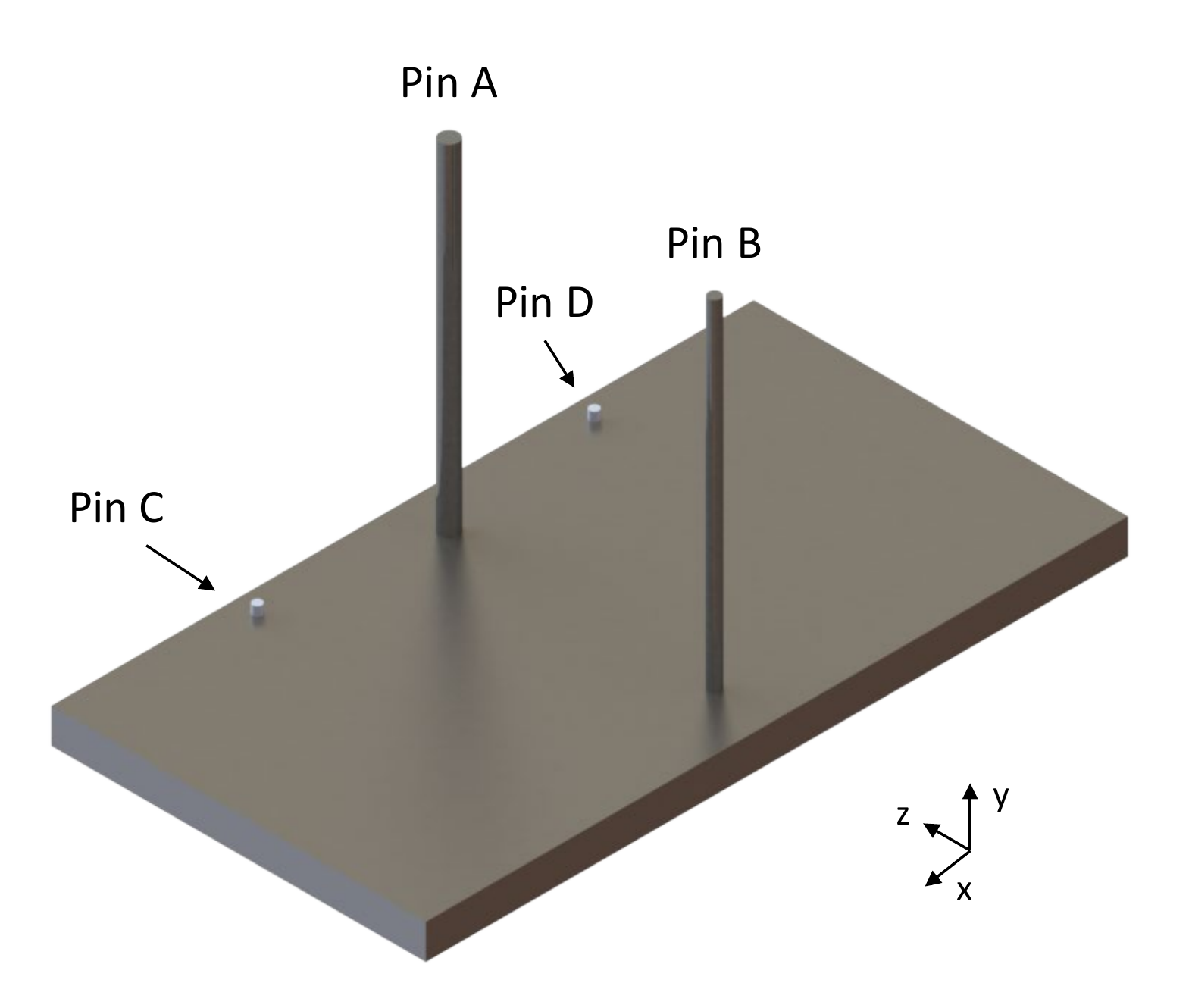}
\vspace{2ex}
\caption{The anticipated alignment methodology for the two grating stacks to form a single grating module \cite{OMeara:2019aa}. The two stacks would be placed on the bottom surface, then abutted against Pins A \& B and either Pin C or Pin D. The precisely machined bottom plate constrains the \^{y}, pitch, and roll of the grating stacks, while pins constrain the remaining three degrees of freedom -- \^{x}, \^{z}, and yaw. Here, Pins A \& B constrain the \^{x} and yaw alignment and Pins C \& D constrain the \^{z} alignment between grating stacks.}
\label{fig:stack-level-alignment}
\end{figure}

Just as with grating-level alignment, the two stacks must be aligned to each other in all six degrees of freedom. The currently envisioned stack-level alignment methodology utilizes a precisely polished surface and additional constraint pins to align the two stacks relative to one another \cite{OMeara:2019aa}. This system is depicted in Figure \ref{fig:stack-level-alignment}. The polished surface (polished to within $\pm1$ \si{\micro}m) orients the two stacks with respect to one another in \^{y}, pitch, and roll, while the pins constrain the stacks in the three remaining degrees of freedom. The sides of grating stacks will be abutted against Pins A \& B to constrain the two stacks in \^{x} and yaw, and the back of each grating stack will be abutted against either Pin C or Pin D to constrain the two stacks in the \^{z} direction.  A more complete discussion of this alignment method can be found in O'Meara et al. (2019) \cite{OMeara:2019aa}. Each degree of freedom for stack-level alignment (as defined in Figure \ref{fig:stack-level-alignment}) and their impact on the LSF is discussed below.

\begin{itemize}[label={}]
    \item \textbf{\^{x}:} A misalignment in this degree of freedom acts to separate the LSFs formed by each grating stack in the cross-dispersion direction. Additionally, it has a small impact on the width of the LSF in the dispersion direction. Due to a weak dependence on both the dispersion and cross-dispersion extent, the \^{x} alignment tolerance has been set to the standard machine tolerance: $\pm127$ micron ($3\sigma$). This allows the two pins that constrain the two stacks in \^{x} (Pin A \& Pin B) to be placed using standard machining techniques. A misalignment of $\pm127$ micron ($3\sigma$) contributes a dispersion extent of $1.0$ \si{\micro}m FWHM and a cross-dispersion extent of $\sim50-180$ \si{\micro}m HPD with the exact impact depending on the specifics of the relative stack-to-stack misalignment.
    \item \textbf{\^{y}:} A misalignment in this degree of freedom will separate the LSFs from each stack in the cross-dispersion direction. Because the wedged substrates are manufactured with a \^{y} tolerance of $\pm127$ \si{\micro}m ($3\sigma$), the maximum misalignment between the two stacks is $\pm254$ \si{\micro}m ($3\sigma$). A small error will also be introduced from the grating module base itself. However, this error ($\pm1$ \si{\micro}m) is negligible compared to the $\pm254$\,\si{\micro}m\,($3\sigma$) uncertainty from wedged substrate manufacture. Therefore, the alignment tolerance in this degree of freedom will be $\pm254$\,\si{\micro}m\,($3\sigma$). This \^{y} misalignment will increase the total extent of the LSF in the cross-dispersion direction by $\sim430-552$ \si{\micro}m HPD.
    \item \textbf{\^{z}:} A misalignment in this degree of freedom will change the dispersion between the two stacks slightly such that the LSFs from each grating stack are dispersed to slightly different locations in the dispersion direction. This will increase the total extent of the combined LSF in the dispersion direction. This alignment will be constrained by pins (Pins C \& D in Figure \ref{fig:stack-level-alignment}). Therefore, to ensure that this constraint can be placed using standard machining, a standard machine tolerance has been adopted in this degree of freedom: $\pm127$\,\si{\micro}m\,($3\sigma$). A misalignment at this level will increase the extent of the LSF in the dispersion direction by $\sim3.6-6.7$\,\si{\micro}m\,FWHM, while only slightly impacting the extent in the cross-dispersion direction. 
    \item \textbf{Pitch (rotation about \^{x}):} A pitch of one grating stack (Grating Stack \#1) relative to the other grating stack (Grating Stack \#2) will cause the LSF formed by Grating Stack \#1 to separate from the LSF formed by Grating Stack \#2 in the cross-dispersion direction. This will increase the extent of the combined LSF formed by the two grating stacks in the cross-dispersion direction. The relative pitch between stacks is governed by both the bottom grating in each grating stack and the flatness of the polished base. With a pitch tolerance on each grating substrate of $\pm30$ arcsec ($3\sigma$), the maximum misalignment between grating stacks induced by the stacks themselves is $\pm60$ arcsec ($3\sigma$). O'Meara et al. (2019) show that the expected worst-case misalignment induced by the polished base is $\pm12.7$ arcsec per grating stack. Thus, a total base-induced misalignment between the two grating stacks could be up to $25.4$ arcsec\cite{OMeara:2019aa}. However, O'Meara et al. (2019) argue that the scenario assumed for this analysis is highly improbable and the likely base-induced misalignment is far below this value. Thus, it is assumed in this error budget that the grating-induced stack misalignment dominates the stack-level pitch misalignment, so an alignment tolerance of $\pm60$\,arcsec\,($3\sigma$) has been adopted for this error. A pitch misalignment at this level will cause the combined LSF formed by the two grating stacks to increase in the cross-dispersion direction by $\sim1800-1910$ \si{\micro}m HPD. 
    \item \textbf{Yaw (rotation about \^{y}):} Similar to a pitch misalignment between the two grating stacks, a yaw misalignment between the grating stacks will cause their individual LSFs to separate in the cross-dispersion direction. This separation will then cause the extent of the combined LSF in the cross-dispersion direction to grow. To limit the extent of the combined LSF in the cross-dispersion direction, this alignment tolerance has been set to $\pm60$ arcsec ($3\sigma$). This leads to a $\sim1020-1130$ \si{\micro}m HPD growth in the cross-dispersion extent of the combined LSF. O'Meara et al. (2019) show that this alignment tolerance is achievable by abutting the two grating stacks against a set of precision pins (Pins A \& B in Figure \ref{fig:stack-level-alignment}) \cite{OMeara:2019aa}.
    \item \textbf{Roll (rotation about \^{z}):} The misalignment of the two grating stacks in roll will cause the individual LSFs (which are much narrower in the dispersion direction when compared to the cross-dispersion direction) to tilt with respect to the mean dispersion direction of the two grating stacks. This will cause the combined LSF to have a greater extent in the dispersion direction. As with stack-level pitch, the stack-to-stack roll alignment is constrained by both the bottom grating in the grating stack and the polished base of the grating module. The grating substrates themselves have a roll requirement of $\pm15$\,arcsec ($3\sigma$), which results in a maximum stack misalignment of $\pm30$ arcsec ($3\sigma$). Additionally, the base can contribute a misalignment of $12.6$ arcsec per stack for a total base-induced roll misalignment of $25.2$~arcsec as derived in O'Meara et al. (2019)\cite{OMeara:2019aa}. However, just as with stack-level pitch, this base-induced misalignment is highly improbable and will likely be much lower than this value. Thus, a $\pm30$ arcsec ($3\sigma$) has been adopted as the alignment tolerance in roll. This misalignment results in a LSF impact of $26.9-31.0$ \si{\micro}m FWHM in the dispersion direction.
\end{itemize}

\noindent A summary of the derived stack-level tolerances can be seen in Table \ref{tab:grating_contributions}. For stack-to-stack alignment, pitch and yaw misalignments have the largest impact on the extent of the LSF in the cross-dispersion direction, while misalignments in \^{z} and roll have the largest impacts on the extent in the dispersion direction.

\subsubsection{Module Alignment}
With a grating module fully assembled, it then must be aligned to its designated 60 degree optic section. Compared to the errors from grating-level and stack-level misalignments though, errors due to misalignments between the grating module and the optic do not contribute significantly to the observed LSF extent in the dispersion and cross-dispersion directions. Instead, these errors shift the LSF centroid on the focal plane. As long as the important spectral lines remain on the detector, this allows module-level alignment tolerances to be slightly looser when compared to grating-level and stack-level alignment tolerances as shown in Table \ref{tab:grating_contributions}.

% Furthermore, the contributions due to module-to-optic misalignments move the entire LSF on the focal plane, but do not affect its total extent in the dispersion and cross-dispersion directions. 

% Since module-to-optic alignment does not impact the extent of the LSF to any large degree, the goal of assigning alignment tolerances for this error term is just to keep the LSF on the detector. 

Misalignments in \^{x}, \^{y}, pitch, and yaw all shift the LSF centroid in the cross-dispersion direction. A movement of $<\pm1$ mm in the cross-dispersion direction has been adopted as the maximum movement that the LSF centroid can be shifted for each of these misalignments.  The size of the detector is $\sim\:$25 x 25 \si{\milli\metre^2}, so even if all four misalignments contributed maximially, the movement of the LSF centroid would be limited to $\sim4$ mm. To limit the movement in this dimension to $\pm1$ mm, \^{x} and \^{y} misalignments have each been given a tolerance of $\pm500$ \si{\micro}m ($3\sigma$). They contribute insignificantly to the extent of the LSF in the dispersion and cross-dispersion directions. Similarly, pitch and yaw misalignments each have a tolerance of $\pm60$ arcsec ($3\sigma$) to limit their cross-dispersion movements to $<\pm1$ mm. Pitch and yaw misalignments at this level also do not contribute significantly to the LSF extent in either the dispersion or cross-dispersion directions.

A misalignment in roll contributes directly to a movement of the LSF centroid in the dispersion direction, since the dispersion direction will change with the roll of the grating module. The goal when assigning an alignment tolerance to this degree of freedom is to keep important lines in the soft X-ray spectrum of Capella on the detector. To satisfy this requirement, the roll tolerance of the grating module relative to the optic has been set to $\pm120$ arcsec ($3\sigma$). 

The remaining misalignment to be considered is a \^{z} misalignment. At the millimeter level, this misalignment only changes the dispersion relation (Eq. 2) on the focal plane. A large change in dispersion on the focal plane will move important lines off the focal plane. A $\sim1$ mm movement of the grating module relative to the optic, however, does not have any appreciable change in the dispersion on the focal plane. Therefore, the alignment tolerance for this degree of freedom is somewhat arbitrary. The \^{z} alignment tolerance has been set to $\pm1$ mm ($3\sigma$) -- a value which should be achievable through standard machine tolerances (even with the stack-up of several interfaces manufactured to standard machine tolerances). A summary of the module-to-optic misalignments and their impact on the observed LSF extent can be seen in Table \ref{tab:grating_contributions}.

\subsection{Forward Assembly Contributions} \label{sect:forward_assembly}

% \begin{table}
% \centering
% \caption{Errors induced by a misalignment of the forward assembly (aligned OGRE mirror module + OGRE grating module) into the observed LSF of the OGRE spectrometer. Shown are the $3\sigma$ level (99.7\%) requirements for each error in all six degrees of freedom (DoF) and the impact of the error in both the dispersion direction (measured as a full-width at half maximum [FWHM]) and the cross-dispersion direction (measured as a half-power diameter [HPD]). }
% \includegraphics[width=\linewidth]{optic-assembly-errors-rev2.pdf}
% \label{tab:optical_assembly_contributions}
% \vspace{-3ex}
% \end{table}

\begin{figure}
\centering
\includegraphics[width=0.99\linewidth]{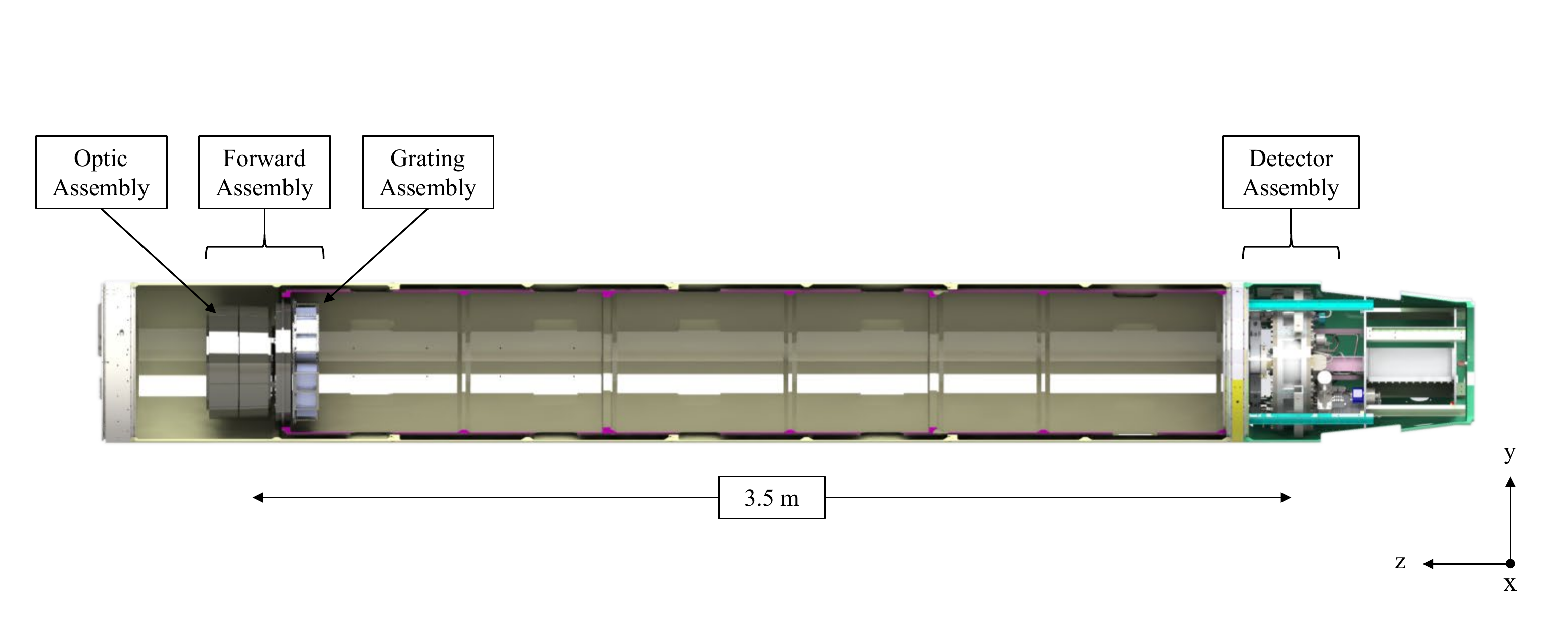}
\caption{CAD rendering of the OGRE payload showing the large separation (3.5 m) between the forward assembly (comprised of the optic and grating assemblies) and the detector assembly. Shown in the bottom-right corner is the coordinate system referenced in Section \ref{sect:forward_assembly}.}
\label{fig:ogre_payload}
\end{figure}

\begin{table*}
\caption{Errors induced by a misalignment of the forward assembly (aligned OGRE mirror module + OGRE grating module) into the observed LSF of the OGRE spectrometer. Shown are the $3\sigma$ level (99.7\%) requirements for each error in all six degrees of freedom (DoF) and the impact of the error in both the dispersion direction (measured as a full-width at half maximum [FWHM]) and the cross-dispersion direction (measured as a half-power diameter [HPD]). Only the maximum LSF impact values are reported for each error in this table.}
\label{tab:optical_assembly_contributions}
\begin{center}   
\begin{tabular}{p{0.17\textwidth}>{\centering}p{0.1\textwidth}>{\centering}p{0.13\textwidth}>{\centering}p{0.13\textwidth}>{\centering}p{0.15\textwidth}>{\centering\arraybackslash}p{0.15\textwidth}}
\toprule
\multirow{2}{*}{Error}&\multirow{2}{*}{DoF}&\multicolumn{2}{c}{Requirement ($3\sigma$)}&\multicolumn{2}{c}{LSF Impact}\\\cline{3-6}
&&\si{\micro}m ($\pm$)&arcsec ($\pm$)&Disp. [\si{\micro}m]&X-Disp. [\si{\micro}m]\\
\toprule
\multirow{6}{2.5cm}{Forward Assembly To Nominal Focal Plane}&X&600&--&--&--\\
&Y&1000&--&--&--\\
&Z&500&--&14.8&--\\
&Pitch (X)&--&60&--&--\\
&Yaw (Y)&--&90&7.9&--\\
&Roll (Z)&--&1000&--&--\\\hline
\multicolumn{4}{l}{\textbf{RSS Total}}&16.8&--\\
\bottomrule
\end{tabular}
\end{center}
\end{table*}

The aligned grating modules and optic assembly become the ``forward assembly''. This forward assembly must be aligned to the nominal focal plane. Whereas the grating-level, stack-level, and module-to-optic alignment tolerances concern components within close proximity to one another, the forward assembly is $\sim3.5$ m away from the nominal focal plane. This introduces new challenges that must be considered when assigning alignment tolerances to this component. Each degree of freedom (as depicted in Figure \ref{fig:ogre_payload}) for this alignment is discussed below.

\begin{itemize}[label={}]
    \item \textbf{\^{x}:} A misalignment in this degree of freedom moves the forward assembly in the dispersion direction relative to the nominal focal plane. This in turn moves the observed LSF on the spectral detectors in the dispersion direction. To keep important lines in the soft X-ray spectrum of Capella on the detector, the \^{x} alignment tolerance has been set to $\pm600$ \si{\micro}m ($3\sigma$).
    \item \textbf{\^{y}:} A misalignment here causes the forward assembly to move in the cross-dispersion direction relative to the nominal focal plane, which also causes the observed LSF to move in the cross-dispersion direction on the spectral detectors. Similar to other LSF centroid translational errors, an alignment tolerance is chosen to limit the movement of the observed LSF on the focal plane to $<\pm1$ mm. This equates to an alignment tolerance of $\pm1000$ \si{\micro}m ($3\sigma$).
    \item \textbf{\^{z}:} A shift in the forward assembly relative to the nominal focal plane in this dimension moves the observed LSF from its nominal focus position. This shift acts to defocus the spectrometer, broadening the LSF extent in the dispersion direction. To limit the impact on the extent of the LSF in the dispersion direction, while taking into consideration the practicality of aligning two components over a $\sim3.5$ m distance, this alignment tolerance has been set to $\pm500$ \si{\micro}m ($3\sigma$). A misalignment at this level increases the LSF extent in the dispersion direction by $13.5-14.8$ \si{\micro}m FWHM.
    \item \textbf{Pitch (rotation about \^{x}):} A pitch of the forward assembly relative to the nominal focal plane acts to move the LSF in the cross-dispersion direction. To limit the translation in this direction to $<\pm1$ mm, an alignment tolerance of $\pm60$ arcsec ($3\sigma$) has been budgeted to this degree of freedom. 
    \item \textbf{Yaw (rotation about \^{y}):} Similar to a yaw misalignment between the grating module and the optic, a yaw misalignment of the forward assembly relative to the nominal focal plane will move the centroid of the observed LSF in the cross-dispersion direction. This movement was limited to $<\pm1$ mm ($3\sigma$), which corresponds to an alignment tolerance of $\pm90$ arcsec. This has a slight impact on the extent of the LSF in the dispersion direction as well ($\sim4.6-7.9$~\si{\micro}m FWHM).
    \item \textbf{Roll (rotation about \^{z}):} A roll misalignment of the forward assembly relative to the nominal focal plane moves the observed LSF in the dispersion direction. To keep important spectral lines on the detector, this tolerance has been set to $\pm0.28$\si{\degree} ($=\pm1000$ arcsec).
\end{itemize}

\noindent The forward assembly will be mounted onto an optical bench cantilevered off of a mounting point close to the focal plane of the spectrometer. To achieve the derived alignment tolerances (as summarized in Table \ref{tab:optical_assembly_contributions}), an adjustable kinematic mount will be designed as an interface between the forward assembly and the optical bench. This mount would provide the forward assembly with movement in \^{z}, pitch, and yaw -- the degrees of freedom with the tightest alignment tolerances. 

A concern with the alignment of the forward assembly is maintaining this alignment during flight. Typically, this optical bench is made from several cylindrical aluminum sections. However, it is anticipated that this aluminum optical bench cannot maintain the derived alignment tolerances due to potential thermal gradients over the length of the bench which would cause an expansion/contraction of the optical bench. To better constrain the forward assembly orientation relative to the nominal focal plane, a custom rigid optical bench will be investigated. 

% Past suborbital rocket payloads have used optical benches made from carbon fiber \cite{France:2016aa}. 

\subsection{Detector Contributions} \label{sect:det_contrib}
\begin{figure}
\centering
\includegraphics[width=0.98\linewidth]{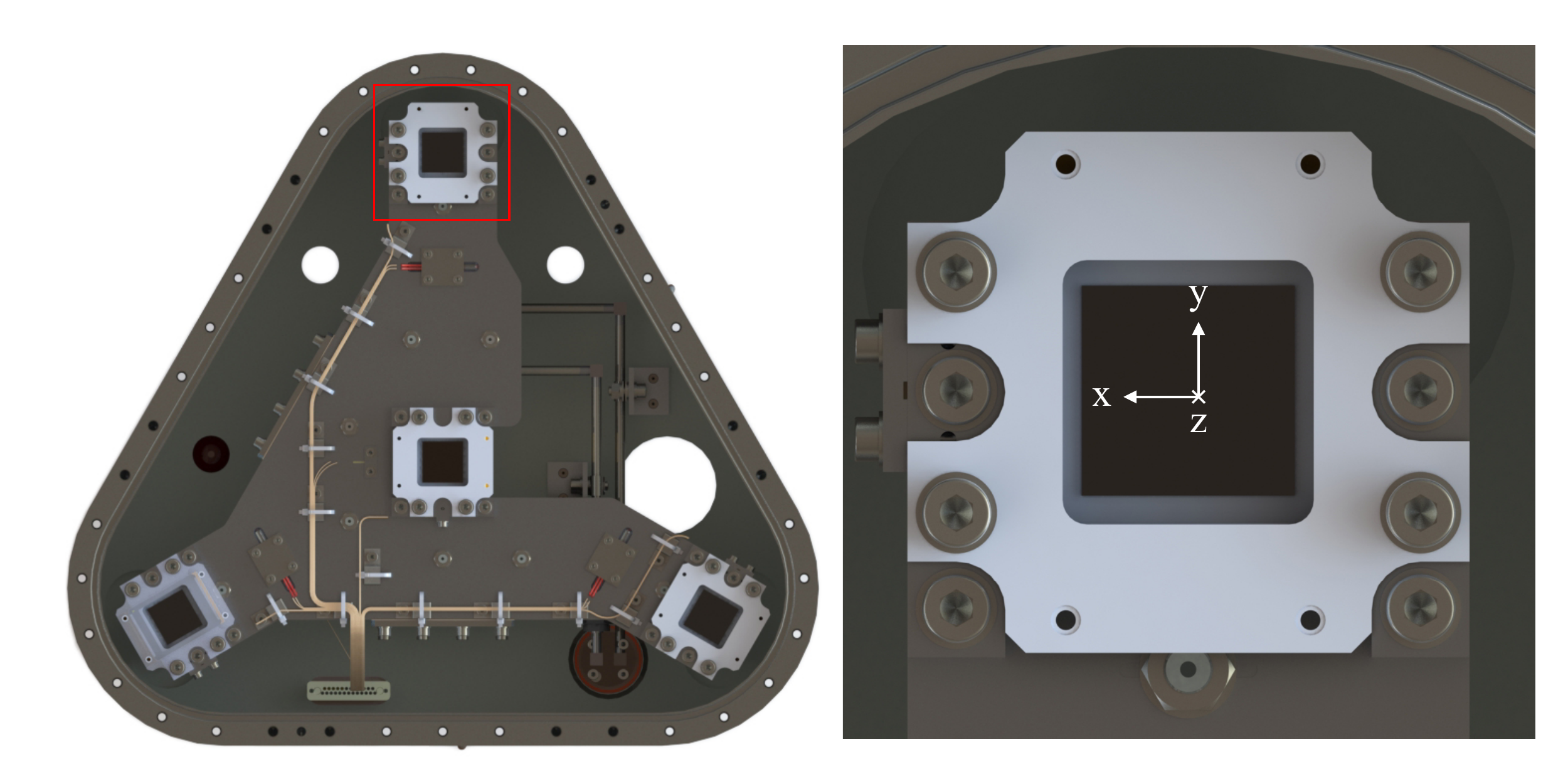}
\caption{A CAD rendering of the OGRE focal plane and a single OGRE spectral detector. \textit{Left} - The OGRE focal plane with three spectral detectors and a central detector. The spectral detectors each sample diffraction arcs from two OGRE grating modules (geometry discussed in Figure \ref{fig:ogre_grating_design}), while the central detector samples the small fraction of light that does not interact with the six OGRE grating modules. \textit{Right} - A zoomed-in view of an OGRE spectral detector (outlined by the red box in the left-hand image). Shown in this image is the coordinate system for this particular spectral detector.} 
\label{fig:ogre_detector}
\end{figure}

The last misalignment in the OGRE spectrometer to consider is the alignment of the spectral detectors relative to the nominal spectrometer focal plane. Just as with all other potential misalignments, this component can be misaligned in all six degrees of freedom. A discussion of each degree of freedom (as depicted in Figure \ref{fig:ogre_detector}) and the impact of a misalignment in each of these degrees of freedom is discussed below.

% \begin{table}
% \centering
% \caption{Errors induced in the observed LSF of the OGRE spectrometer by a misalignment of a spectral detector relative to the nominal focal plane . Shown are the $3\sigma$ level (99.7\%) requirements for each error in all six degrees of freedom (DoF) and the impact of the error in both the dispersion direction (measured as a full-width at half maximum [FWHM]) and the cross-dispersion direction (measured as a half-power diameter [HPD]). }
% \includegraphics[width=\linewidth]{detector-errors-rev2.pdf}
% \label{tab:detector_contributions}
% \vspace{-3ex}
% \end{table}

\begin{table*}
\caption{Errors introduced into the observed LSF of the OGRE spectrometer by a misalignment of a spectral detector relative to the nominal focal plane . Shown are the $3\sigma$ level (99.7\%) requirements for each error in all six degrees of freedom (DoF) and the impact of the error in both the dispersion direction (measured as a full-width at half maximum [FWHM]) and the cross-dispersion direction (measured as a half-power diameter [HPD]). Only the maximum LSF impact values are reported for each error in this table.}
\label{tab:detector_contributions}
\begin{center}   
\begin{tabular}{p{0.17\textwidth}>{\centering}p{0.1\textwidth}>{\centering}p{0.13\textwidth}>{\centering}p{0.13\textwidth}>{\centering}p{0.15\textwidth}>{\centering\arraybackslash}p{0.15\textwidth}}
\toprule
\multirow{2}{*}{Error}&\multirow{2}{*}{DoF}&\multicolumn{2}{c}{Requirement ($3\sigma$)}&\multicolumn{2}{c}{LSF Impact}\\\cline{3-6}
&&\si{\micro}m ($\pm$)&arcsec ($\pm$)&Disp. [\si{\micro}m]&X-Disp. [\si{\micro}m]\\
\toprule
\multirow{6}{2.5cm}{Detector To Nominal Focal Plane}&X&600&--&--&--\\
&Y&254&--&--&--\\
&Z&254&--&8.0&--\\
&Pitch (X)&--&3600&--&--\\
&Yaw (Y)&--&3600&--&--\\
&Roll (Z)&--&3600&--&--\\\hline
\multicolumn{4}{l}{\textbf{RSS Total}}&8.0&--\\
\bottomrule
\end{tabular}
\end{center}
\end{table*}

\begin{itemize}[label={}]
    \item \textbf{\^{x}:} A misalignment in this degree of freedom causes a shift of the detector in the dispersion direction relative to the nominal focal plane. If shifted too far, important spectral lines from Capella will begin to fall off of the detector. To keep important lines on the detector, an alignment tolerance of $\pm600$ \si{\micro}m has been assigned to this degree of freedom. 
    \item \textbf{\^{y}:} A misalignment here causes the detector to move in the cross-dispersion direction relative to the nominal focal plane. As a result of this movement, the observed LSF will move on the detector. However, the observed LSF just needs to remain on the detector. An alignment tolerance of $\pm254$ \si{\micro}m has been assigned here, but realistically this could be loosened if necessary.
    \item \textbf{\^{z}:} A shift in \^{z} of the detector relative to the nominal focal plane moves the observed LSF away from its nominal focus position. A misalignment here then acts to defocus the spectrometer which introduces aberrations into the LSF in the dispersion direction. To limit the impact on the extent of the observed LSF in the dispersion direction, this alignment tolerance has been set to $\pm254$ \si{\micro}m. A misalignment at this level increases the LSF extent in the dispersion direction by $7.2-8.0$ \si{\micro}m FWHM.
    \item \textbf{Pitch (rotation about \^{x}):} A pitch of the detector (about its center) relative to the nominal focal plane will act to defocus the LSF -- one half of the LSF will be intrafocal and the other half will be extrafocal. However, only relatively large pitches ($>2$\si{\degree}) will cause any appreciable impact on the extent of the observed LSF. Therefore, a relatively loose tolerance of $\pm1$\si{\degree} has been adopted here. This tolerance can be loosened if necessary.
    \item \textbf{Yaw (rotation about \^{y}):} Similar to a misalignment of the detector in pitch, a yaw misalignment of the detector (about its center) relative to the nominal focal plane will move the portions of the spectrum on the detector out of its nominal focal position. Similar to pitch though, only large yaws ($>2$\si{\degree}) will lead to any appreciable impact on the observed LSF. Therefore, the same $\pm1$\si{\degree} alignment tolerance has been adopted here, but this could be loosened if necessary.
    \item \textbf{Roll (rotation about \^{z}):} A roll of the detector relative to the nominal focal plane will have no impact on the observed LSF. Its alignment tolerance has been set to $\pm1$\si{\degree}$=3600$ arcsec, but this tolerance is somewhat arbitrary. 
\end{itemize}

\noindent As shown in Table \ref{tab:detector_contributions}, detector contributions only impact the extent of the observed LSF at the $\sim8$~\si{\micro}m level. Compared to contributions from the grating and the optic, misalignments of the detector do not impact the LSF extent to any appreciable degree, only serving to move the LSF centroid on the detector.

One complication in the alignment of the detectors is that their placement within their packaging from the manufacturer (e2v) is not known to the level of the derived alignment tolerances. Therefore, their placement must be reconstructed once assembled and then adjusted to move them into their optimal location within the alignment tolerances. Since the detectors cannot be touched by traditional measuring devices such as a portable measuring arm or a coordinate-measuring machine, a non-contact measuring method is required. The baseline method for this non-contact measurement is a laser scanner attachment to a portable measuring arm. With an accuracy of $\sim40$ \si{\micro}m, this laser scanner would be able to measure each detector within its packaging to the required accuracy. With its position and orientation known with respect to its packaging, each detector could then be adjusted to place it in its proper location to within the required alignment tolerances. 

\subsection{In-Flight Contributions: Jitter}
The final contributor to the LSF of the OGRE spectrometer is the in-flight contribution from jitter -- the pointing stability of the payload during flight. The frequency of this movement is expected to be higher than the readout cadence of the detector, so the jitter of the payload about the source will artificially increase the size of the source. This therefore impacts the extent of the observed LSF in both the dispersion and cross-dispersion directions.

The NASA Sounding Rockets User Handbook \cite{NASASRHB} states that the NSROC (NASA Sounding Rocket Operations Contract) Celestial Attitude Control System can achieve a jitter of $<1$ arcsec/sec FWHM in its linear thrust configuration (the configuration baselined for the OGRE mission). This number has also been confirmed from past sounding rocket flight data. Therefore, this number has been adopted as the jitter requirement for the OGRE spectrometer. This requirement corresponds to an induced aberration of 17 \si{\micro}m FWHM in the dispersion direction and 14.4 \si{\micro}m HPD in the cross-dispersion direction of the LSF.

\section{Combining Errors}
With all performance requirements and alignment tolerances derived, they can be combined to inform the achievable performance of the OGRE spectrometer. Errors were first analyzed as if they were independent and all contributed at their $3\sigma$ values, then raytrace simulations of the OGRE spectrometer with these same errors and misalignments were performed to compare results. Misalignments were then randomized within their $3\sigma$ limits and further raytrace simulations were performed to understand spectrometer performance when misalignments do not all contribute at their $3\sigma$ limits.

% \begin{table}
% \centering
% \caption{All errors contributing to the observed LSF of the OGRE spectrometer, including errors from the optic assembly, grating module, forward assembly (aligned optic + grating module), detector, and in-flight contributions. Listed is each error and its contribution to the observed LSF in both the dispersion direction (measured as a full-width at half maximum [FWHM]) and the cross-dispersion direction (measured as a half-power diameter [HPD]). }
% \includegraphics[width=\linewidth]{total-errors-rev2.pdf}
% \label{tab:all_contributions}
% \vspace{-3ex}
% \end{table}

\begin{table*}
\caption{All errors contributing to the observed LSF of the OGRE spectrometer, including errors from the optic assembly, grating module, forward assembly (aligned optic + grating module), detector, and in-flight contributions. Listed is each error and its contribution to the observed LSF in both the dispersion direction (measured as a full-width at half maximum [FWHM]) and the cross-dispersion direction (measured as a half-power diameter [HPD]). }
\label{tab:all_contributions}
\begin{center}   
\begin{tabular}{p{0.25\textwidth}>{\centering}p{0.15\textwidth}>{\centering\arraybackslash}p{0.15\textwidth}}
\toprule
\multirow{2}{*}{Component}&\multicolumn{2}{c}{LSF Impact}\\\cline{2-3}
&Disp. [\si{\micro}m]&X-Disp. [\si{\micro}m]\\
\toprule
Optic&25.4&84.8\\
Grating&41.2&2367.9\\
Forward Assembly&16.8&--\\
Detector&8.0&--\\
Jitter&17.0&14.4\\\hline
\textbf{RSS Total}&54.6&2369.5\\
\bottomrule
\end{tabular}
\end{center}
\end{table*}

\subsection{The RSS Method: Independent Errors}
If each induced error to the LSF is asssumed to be independent with respect to the other induced errors, LSF errors add in quadrature (root sum of the squares; RSS). In Tables 1-4, the individual errors from each component were added in quadrature with their total LSF impact (``RSS Total'') in both the dispersion and cross-dispersion directions displayed at the bottom of each table. These totals were collected and are displayed together in Table \ref{tab:all_contributions}. The totals from each component were then added in quadrature to yield a total LSF impact in both the dispersion and cross-dispersion directions. If all errors contribute maximally, the resulting total impact on the observed LSF is 54.6 \si{\micro}m FWHM in the dispersion direction and $\sim2370$ \si{\micro}m HPD in the cross-dispersion direction. Converting the extent of the LSF in the dispersion direction ($\Delta x$) to spectral resolution ($R=x/\Delta x$ with $x=98.2$ mm), the maximum achievable spectral resolution if all LSF errors were at their $3\sigma$ limit and added in quadrature is $R\approx1800$. This result is below the spectral resolution goal of $R>2000$, but it still comfortably meets the resolution requirement of $R>1500$. 

\subsection{Raytrace Simulation Results}
\begin{figure}
\begin{center}
\begin{tabular}{c}
\includegraphics[width=0.97\linewidth]{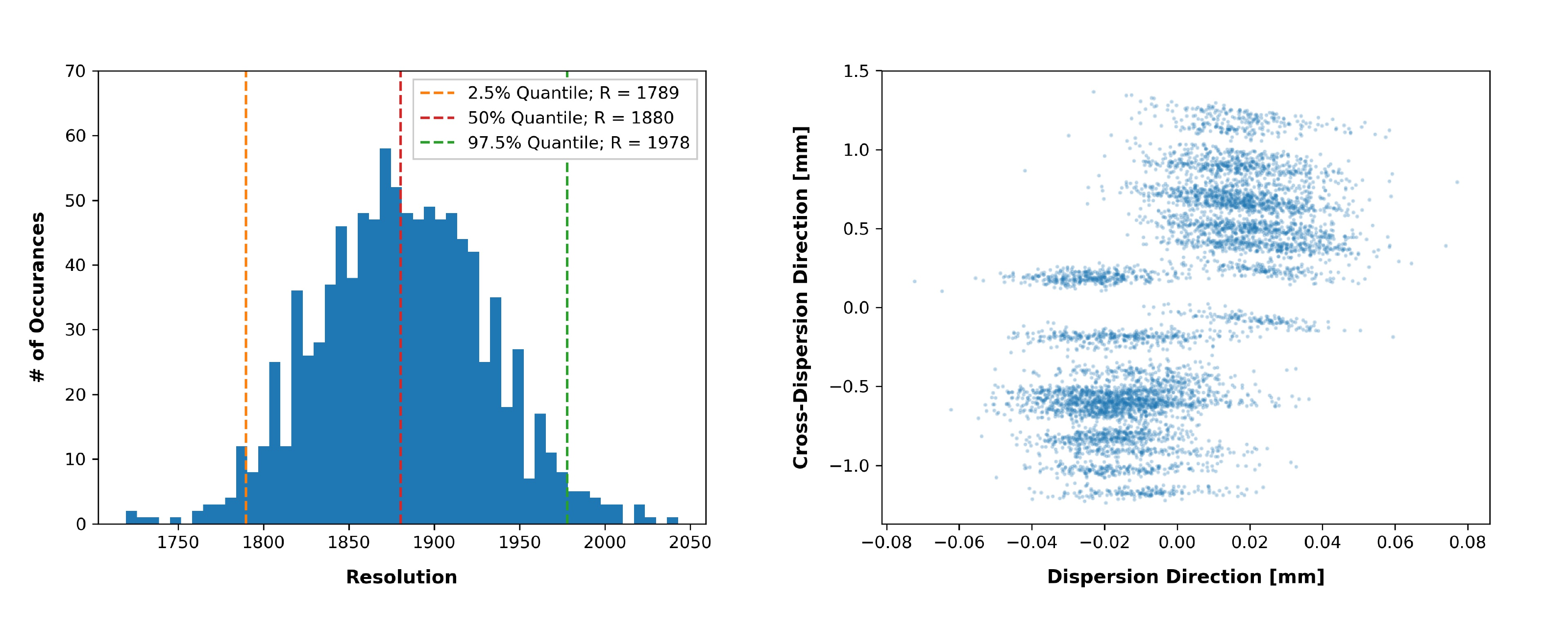}  % fig2 includes two images 
\\
(a) \hspace{0.45\linewidth} (b)
\end{tabular}
\end{center}
\caption 
{ \label{fig:max_res_distribution}
Results from 1,000 raytrace simulations of the OGRE spectrometer with misalignments, performance errors, and jitter at their $3\sigma$ limits (presented in Tables 1-4). (a): Distribution of the achieved spectral resolutions from the 1,000 raytrace simulations. These values range from $R\sim1790$ ($2\sigma$ low) to $R\sim1980$ ($2\sigma$ high), with a median spectral resolution of $R=1880$. (b): The diffracted-order LSF from a single raytrace simulation in (a). Structure within this LSF comes from the two misaligned grating stacks (two groups centered at $x\sim-0.02$ and $x\sim0.02$ in the dispersion direction) and also from the individual gratings within each stack (quasi-horizontal lines within the two grating stack groups). This diffracted-order LSF achieves a spectral resolution of $R\sim1880$.} 
\end{figure}

To test whether the assumption that all errors are independent and add in quadrature is a fair assumption, raytrace simulations of the OGRE spectrometer were performed with all misalignments and errors at their $3\sigma$ limits (the $3\sigma$ requirements presented in Tables 1-4). A total of $N_{sim}=1000$ simulations of the OGRE spectrometer were performed with these misalignments and errors at the blaze wavelength of the system: $n\lambda=4.76$ nm. For each simulation, the spectral resolution at the blaze wavelength was calculated. The distribution of the resulting resolution values attained from simulations is presented in Figure \ref{fig:max_res_distribution}a. Further, the result of an individual simulation is shown in Figure \ref{fig:max_res_distribution}b. The results from this raytrace simulation are consistent with the results obtained when assuming all errors are independent and add in quadrature -- $R\approx1880$ (median value). So therefore, if all errors contributed at their $3\sigma$ error limits, the OGRE spectrometer would not achieve its spectral resolution goal of $R>2000$, but would still comfortably achieve its spectral resolution requirement of $R>1500$ at this wavelength.

It should be noted that diffracted-order LSF resulting from a single simulation of the OGRE spectrometer in Figure \ref{fig:max_res_distribution}b demonstrates structure; it is apparent that two separate LSFs make up this combined LSF. This structure suggest that the LSFs formed by the individual grating stacks could potentially be used to provide a higher resolution spectra, since the individual LSFs are much narrower than the combined LSF. Identification and subsequent analysis of these individual LSFs that make up the combined LSF would require thorough calibration of the instrument prior to launch to ensure all four LSFs on a single spectral detector -- two from this grating module and another two from a second grating module clocked 60 degrees relative to the first (geometry shown in Figure \ref{fig:ogre_grating_design}) -- can be individiually identified. The potential for this analysis will be investigated in the future.

% individual spectrometer components performing at their maximum permissible error levels, misalignment of each component at the maximum value allowed by their alignment tolerances, and jitter consistent with the expected performance of the attitude control system

\begin{figure}
\centering
\includegraphics[width=0.49\linewidth]{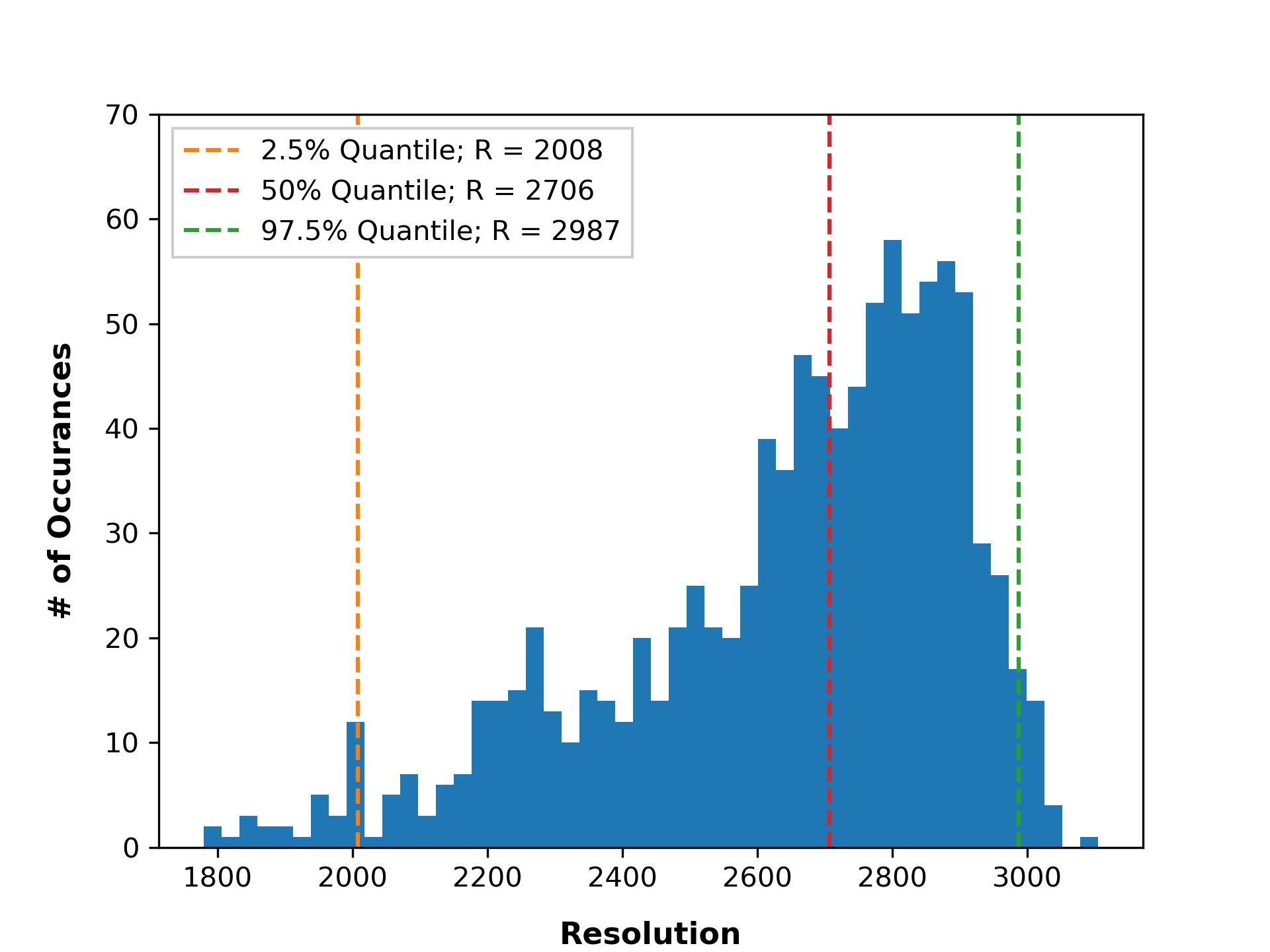}
\vspace{2ex}
\caption{Results from 1,000 raytrace simulations of the OGRE spectrometer with misalignment values chosen randomly within their $3\sigma$ requirements, but with performance contributions (optic performance, grating period errors, and jitter) still contributing at their $3\sigma$ error limits. The achieved spectral resolutions range from $R\sim2000$ ($2\sigma$ low) to $R\sim2990$ ($2\sigma$ high), with a median spectral resolution of $R=2700$.}
\label{fig:random_res_distribution}
\end{figure}

The previous simulations assumed all misalignments and performance errors contributed at their $3\sigma$ error limit. In reality though, some misalignments will be well within their $3\sigma$ limit, while others might be closer to the periphery of their limits. To understand how this situation might improve the achievable resolution of the OGRE spectrometer, further raytrace simulations were performed. However, instead of assuming that all contributions to the LSF were at their $3\sigma$ limits, all misalignments were given random values within their $3\sigma$ parameter space. Performance contributions (optic performance, grating resolution limit, and jitter) were kept at their $3\sigma$ limits, since values for these contributions are expected to be closer to their $3\sigma$ limits. Similar to the previous simulations, this raytrace simulation was performed $N_{sim}=1000$ times with each simulation having unique misalignment values. The results of these simulations are presented in Figure \ref{fig:random_res_distribution}. These simulations show that while there are still some incarnations of the spectrometer that do not achieve the $R>2000$ performance goal ($<2.5$\% of simulations), the vast majority of spectrometers modelled achieved a spectral resolution beyond this goal ($>97.5$\%). 

\section{SUMMARY}

In this manuscript, a comprehensive LSF error budget for the soft X-ray grating spectrometer on the Off-plane Grating Rocket Experiment (OGRE) was described. This error budget described potential impacts to the LSF observed by the OGRE spectrometer including component misalignments and performance errors, derived realistic alignment tolerances for component misalignments, and elucidated how each misalignment and performance error impacted the observed LSF. The impacts from component misalignments and performance errors were combined first by assuming independent errors such that the errors added in quadrature. It was found that if all errors contribute at their $3\sigma$ error limits, the spectrometer would achieve a spectral resolution of $\sim1800-1900$. This performance meets the OGRE spectral resolution requirement of $R>1500$ comfortably, but does not meet the spectral resolution goal of $R>2000$. However, if not all misalignments contributed at their $3\sigma$ error limits, but instead were randomly distributed within their $\pm3\sigma$ error limits, the spectrometer meets the spectral resolution goal of $R>2000$ for $>97.5$\% of the simulations. These results suggest that the OGRE spectrometer should be able to comfortably meet its spectral resolution requirement of $R>1500$, and, depending on the exact values of misalignments during assembly, even meet its spectral resolution goal of $R>2000$.

% \disclosures 
\subsection*{Disclosures}
The authors have no relevant financial interests in the manuscript and no other potential conflicts of interest to disclose.

\acknowledgments 
The work presented in this manuscript is supported by NASA grant NNX17AD19G, a NASA Space Technology Research Fellowship, and internal funding from The Pennsylvania State University. The authors would like to thank current and past members of the McEntaffer research group for their support of this project. This work makes use of PyXFocus, an open-source Python-based raytrace package.  

%%%%% References %%%%%

\bibliography{report}   % bibliography data in report.bib
\bibliographystyle{spiejour}   % makes bibtex use spiejour.bst

%%%%% Biographies of authors %%%%%

\vspace{2ex}\noindent\textbf{Benjamin D. Donovan} is a PhD candidate at The Pennsylvania State University in the Department of Astronomy \& Astrophysics. He received his B.S. in Physics \& Astronomy from The University of Iowa in 2016, and his M.S. in Astronomy \& Astrophysics from The Pennsylvania State University in 2019. He is the recipient of a NASA Space Technology Research Fellowship. His current research interests include the design and development of space-based astronomical instrumentation. He is a member of SPIE.

\vspace{1ex}
\noindent Biographies and photographs of the other authors are not available.

\listoffigures
\listoftables

\end{spacing}
\end{document}